\documentclass[prd,aps,amssymb,nofootinbib,showpacs,preprint]{revtex4}
\usepackage{amsmath,amsfonts}
\usepackage{textcomp,color}
\usepackage{hyperref}
\usepackage{graphics,graphicx}
\usepackage{epstopdf}
\usepackage{ulem}
\usepackage{overpic}

\begin{document}
	\title{Multidimensional gravity with higher derivatives and inflation}
	\author{J\'ulio C. Fabris}
	\affiliation{N\'ucleo Cosmo-ufes \& Departamento de F\'{\i}sica, Universidade Federal do Esp\'{\i}rito Santo, Vit\'oria, ES, CEP 29075-910, Brazil}
	\affiliation{National Research Nuclear University MEPhI (Moscow Engineering Physics Institute), 115409, Kashirskoe shosse 31, Moscow, Russia}
	\author{Arkady~A.~Popov}
	\affiliation{N.~I.~Lobachevsky Institute of Mathematics and Mechanics, Kazan  Federal  University, \\ 420008,   
		Kremlevskaya  street  18,  Kazan,  Russia}
	\author{Sergey~G.~Rubin}
	\affiliation{National Research Nuclear University MEPhI (Moscow Engineering Physics Institute),\\ 115409, Kashirskoe shosse 31, Moscow, Russia}
	\affiliation{N.~I.~Lobachevsky Institute of Mathematics and Mechanics, Kazan  Federal  University, \\ 420008,   
		Kremlevskaya  street  18,  Kazan,  Russia}
	
	\begin{abstract}
		We elaborate on the inflationary model starting from multidimensional Lagrangian and gravity with second-order curvature terms. The effective scalar field is related to the Ricci scalar of extra dimensions. It is shown that the Kretschmann and the Ricci tensor square terms dominate during inflation.  The observable values of the spectral index and the tensor-to-scalar ratio are obtained for specific values of the model parameters.
		%Another widespread idea, the extra-dimensional world is considered as the necessary element for a future theory. The paradigm of the extra dimensions is also attracted to the explanation of the cosmological evolution.  The simplest way to explain their invisibility for modern experiments is an imposing a compact extra space with the size smaller than $10^{-18}$cm. 
		
	\end{abstract}
	
	\maketitle
	% * <sergeirubin@list.ru> 2018-08-24T21:38:25.596Z:
	%
	% ^.
	
	\section{Introduction}
	
	It is usually assumed that our Universe was nucleated at the Planck energies and evolved by expanding and cooling to the present condition. The inflationary stage is characterized by sub-Planckian energy density and looks unavoidable. Spontaneous creation of the Universe with the inflationary regime has been elaborated in \cite{Firouzjahi:2004mx}.
	Models describing the inflation have been elaborated using many different ingredients, like supersymmetry  \cite{Antusch:2011wu} and attracting the inflationary idea for an explanation of other cosmological problems like baryogenesis \cite{1997PhRvD..56.6155D}  and primordial black holes \cite{2005APh....23..265K}, \cite{2009NuPhB.807..229D}. For review, see \cite{Cline:2018fuq,Akrami:2018odb} and references therein.
	%Due to this Grand Unified Theory, in its different formulations, has been studied, initially as the fundamental framework for inflation, and some success and with some drawbacks. However, GUT remains a speculative theoretical proposal. 
	On the other hand, as the energy scale is high enough some effects related to a quantum regime may be manifested and be responsible for the inflationary regime. Two typical elements from quantum gravity may be relevant in the construction of inflationary models: non-linear geometrical extensions of General Relativity and extra dimensions. Moreover, the quantization of gravity asks for non-linear geometric extension of the Einstein-Hilbert action. The first and most successful formulation of the inflationary model, the Starobinsky model \cite{Starobinsky:1980te}, considers non-linear geometric terms belonging to the $f(R)$ class of theories. 
	
	The gravity with higher derivatives is widely used in modern research despite the internal problems inherent in this approach \cite{1988PhLB..214..515B}, \cite{2015arXiv150602210W}. Attempts to avoid the Ostrogradsky instabilities are made \cite{2017PhRvD..96d4035P} and extensions of the Einstein-Hilbert action attract much attention.  The $f(R)$-gravity is one of the simplest extension of the Einstein-Hilbert gravity and it is free of Ostrogradsky instabilities. Reviews \cite{DeFelice:2010aj}, \cite{2011PhR...509..167C} contain description of the $f(R)$-theories including extension to the Gauss-Bonnet gravity. Examples of research with a specific form of the function $f(R)$ can be found in
	\cite{2006GrCo...12..253S}, \cite{2006A&AT...25..447S}. Most of the research assumes positive curvature of extra space metric.
	
	Another widespread idea, the extra-dimensional world is considered as the necessary element for a fundamental complete theory. The paradigm of the extra dimensions is also attracted for an explanation of the cosmological evolution \cite{Abbott:1984ba}.  The simplest way to explain their invisibility is an imposing a compact extra space with the size smaller than $10^{-18}$cm. 
	
	Our research aims to unify two necessary elements of a future theory - the (compact) extra dimensions and the gravity with higher derivatives to build a model of inflation. No matter fields are assumed from the beginning.  It is shown that the Kretschmann and the Ricci tensor square terms dominate during inflation and hence play a significant role in our research.

	The inflation gives information which should be reproduced by our scenario. More definitely, it is assumed that the effective potential should provide us with the reheating stage and appropriate values of the scalar spectral index $n_s$ and the scalar/tensorial relative amplitude $r$. In the present paper, we investigate the conditions to have a successful inflationary model from a non-linear theory of gravity in higher dimensions.
	
	Throughout this paper we use the conventions for the curvature tensor $R_{ABC}^D=\partial_C\Gamma_{AB}^D-\partial_B\Gamma_{AC}^D+\Gamma_{EC}^D\Gamma_{BA}^E-\Gamma_{EB}^D\Gamma_{AC}^E$
	and for the Ricci tensor $R_{MN}=R^F_{MFN}$.

	\section{Setup}
	In this paper we consider a $D=4+n$ - dimensional manifold $\mathbb{M}$, having the simplest geometric structure of a direct product, $\mathbb{M}=\mathbb{M}_4 \times \mathbb{M}_n$, with the metric
	\begin{equation}\label{metric}
	ds^2 = dt^2 - a(t)^2\Big(dr^2 +r^2(d\theta^2 +\sin^2(\theta)d\varphi^2 \Big)- e^{2\beta(t)} d\Omega_n^2,
	\end{equation}
	Here we assume that extra space is $n$-dim maximally symmetrical one with positive curvature.
	
	It is known that quantum effects inevitably lead to nonlinear corrections in the expression for the action \cite{Donoghue:1994dn}. In this case, the action should contain terms with higher derivatives. We start our study  with action in the following form
	\begin{equation}\label{Lgen}
	S_{gen}=\frac12 m_D^{D-2}\int d^D x \sqrt{g_D}[f(R)+c_1R_{AB}R^{AB}+c_2R_{ABCD}R ^{ABCD}],
	\end{equation}
	where $f(R)$ is an arbitrary function
	of the Ricci scalar $R$, $m_D$ is the $D$-dimensional Planck mass and $c_1, c_2$ are free parameters of
	the Lagrangian. This action can be considered as a basis of an effective theory \cite{2007ARNPS..57..329B}.
	
	It is a priory assumed that the extra space size is small compared to the size of our 4-dim space during inflation so that
	the following inequality for the Ricci scalars
	\begin{equation}\label{ineq0}
	R_{4}\ll R_{n}
	\end{equation}
	holds.  This inequality will be checked at the final stage of the study. In this case, we can follow the method elaborated in \cite{Bronnikov:2005iz, 2007CQGra..24.1261B}.
	Metric \eqref{metric} leads to the Ricci scalar in the form
	\begin{equation}\label{Tailor}
	R=R_{4} + R_{n}+ P_k; \quad P_k = 2n\ddot{\beta} +n(n+1)(\dot{\beta})^2 +6 n\frac{\dot{a}}{a} \dot{\beta}.
	\end{equation}
	{\it The} additional inequality
	\begin{equation}\label{ineq00}
	P_k\ll R_{n}
	\end{equation}
	means that evolution of the metric coefficient $\beta(t)$ is slow.  More specifically,
	\begin{equation}\label{ep}
	|\partial_{\mu} g_{AB}| \sim \epsilon |\partial_a  g_{AB}|,
	\end{equation}                               
	that is, each derivative $\partial_{\mu}$ contains a small parameter $\epsilon \ll 1$.
	
	According to formulas (\ref{metric} -- \ref{ineq00}) we can perform the Taylor decomposition of the function $f(R)$
	\begin{eqnarray}\label{actJ}
	&&S=\frac{1}{2}V_{n}\int d^{d_0}x \sqrt{-g_0}e^{d_2\beta_n} [f'(R_{n})R_{4} + f'(R_{n})P_k + f(R_{n})+ \nonumber \\
	&& +c_1R_{AB}R^{AB}+c_2R_{ABCD}R ^{ABCD}],   \\
	&&   R_n=\phi=n(n-1)e^{-2\beta(t)},  \nonumber \\
	&&R_{AB}R^{AB}= \frac{(n-1)^2}{n}e^{-4\beta}+2n(n-1)e^{-2\beta}(\ddot{\beta}_2 + n\dot{\beta}^2 + 4\frac{\dot{a}}{a}\dot{\beta})+O(\epsilon^4),  \nonumber \\
	&&R_{ABCD}R ^{ABCD}= 2\frac{n-1}{n}e^{-4\beta}+4n(n-1)e^{-2\beta}\dot{\beta}_2^2 +O(\epsilon^4) \nonumber
	\end{eqnarray}
	after some tedious calculations. In the following, the unit $m_D =1$ is used. The Ricci tensor square and the Kretschmann scalar have been also expressed explicitly, see \cite{Bronnikov:2005iz,2013bhce.conf.....B}.

	The action of the form \eqref{actJ} is written in the Jordan picture. We consider this as the "physical" frame which gives us the Planck mass, in particular
	\begin{equation}\label{MPl}
	M_{P}^2=V_{n}e^{n\beta_m}f'(\phi_m);\quad V_{n}=\frac{2\pi^{\frac{n+1}{2}}}{\Gamma(\frac{n+1}{2})}.
	\end{equation}
	Here $\phi_m$ delivers minimum of $V_E(\phi)$ (see definition \eqref{V} below).
	
	It is more familiar to work in the Einstein frame. To this end, we have to perform the conformal transformation
	\begin{equation}\label{conform}
	g_{ab}\rightarrow g_{ab}^{(E)}=e^{n\beta} |f'(\phi)|g_{ab}
	\end{equation}
	of the metric describing the subspace $M_{4}$.
	That leads to the action in the Einstein frame in form
	\begin{equation}\label{Lscalar}
	S_{low}=\frac12 V_n \int d^4 x \sqrt{g_4}\ \mbox{sign}(f')[R_4 + K(\phi)(\partial \phi)^2 -2V(\phi) ],
	\end{equation}
	\begin{eqnarray} \label{K}
	&& K(\phi)=\frac{1}{4\phi^2}\biggl[6\phi^2 (f''/f')^2 - 2n\phi(f''/f')+\frac12 n(n+2)\biggr] +\frac{c_1+c_2}{f'\phi},
	\end{eqnarray}
	\begin{eqnarray} \label{V}
	&&V(\phi)=-\frac{\mbox{sign}(f')}{2f'^2}\biggl[\frac{|\phi|}{n(n-1)}\biggr]^{n/2}\biggl[f(\phi)+\frac{c_V}{n}\phi^2\biggr], \quad c_V=c_1 + \frac{2c_2}{(n-1)}
	\end{eqnarray}
	representing specific Lagrangian of the scalar-tensor gravity \cite{Bronnikov:2005iz}. Here $D=4+n$ and the physical meaning of the effective scalar field $\phi$ is the Ricci scalar of the extra space.
	
	An important remark is necessary. The action \eqref{actJ} describes the field evolution at high energies in the Jordan frame. When the scalar field is settled in its minimum we should express the 4-dim Planck mass according to \eqref{MPl}.
	At the same time, the scalar field is evolving during inflation and it is worth to use the Einstein frame to facilitate analysis. In this case, we should consider
	\begin{equation}\label{m4}
	m_4\equiv\sqrt{V_n}
	\end{equation}
	as the effective Planck mass during the inflation.
	
	It is worth to foresee how this model is related to the present epoch, what implies several conditions. In this stage, the space-time metric is (almost) Minkowskian due to the smallness of the cosmological constant. Necessary conditions are as follows:
	\begin{equation}\label{VMink}
	V(\phi_m)=0;\quad V'(\phi_m)=0.
	\end{equation}
	The inequalities
	\begin{equation}\label{ineq}
	\phi_m >0,\quad f'(\phi_m)>0,\quad K(\phi_m)>0.\quad V''(\phi_m)=m^2 >0;
	\end{equation}
	are necessary to consider the scalar field $\phi$ as the field with the standard properties.
	The function $f(R)$ should be specified to make specific predictions. It is chosen in the form
	\begin{equation}\label{fR}
	f(R)=a R^2 + b R +c.
	\end{equation}
	The parameters $m_D,a,b,c,c_1,c_2$ are not arbitrary. We may renormalize the parameter $m_D$ to equate one of other parameters to unity. It is assumed in section \ref{secpar} that $b=1$.
	
	Keeping in mind quadratic form \eqref{fR} of the function $f$ one can easily solve algebraic equations \eqref{VMink} with the result
	\begin{equation}\label{cond1}
	\phi_m =-\frac{b}{2(a+c_V/n)}
	\end{equation}
	and with the following relation between the Lagrangian parameters
	\begin{equation}\label{cond2}
	c=\frac{b^2}{4(a+c_V/n)}.
	\end{equation}
	Our space-time with FLRW metric is surely not Minkowskian so that the relation \eqref{cond2} holds approximately.
	
	Consider inequalities \eqref{ineq} in more detail. First inequality in \eqref{ineq} gives
	\begin{equation}\label{cond6}
	-\frac{b }{2 \left(a +c_V/n\right)} > 0.
	\end{equation}
	Second inequality in \eqref{ineq} leads to
	\begin{equation}\label{cond3}
	f'(\phi_m)>0\rightarrow \frac{b c_V/n}{ \left(a +c_V/n\right)} > 0.
	\quad \mbox{or} \quad c_V < 0 %sign\left(\frac{c_V b}{2a+c_V}\right) \cdot (2a+c_V)>0 \rightarrow sign (c_V b)<0.
	\end{equation}
	Third inequality in \eqref{ineq} leads to inequality
	\begin{equation}\label{cond4}
	\left[ \left( 12 a^2 +4 a c_V +c_V^2 \right)n +2c_V^2 -4 c_V (c_1 +c_2) \right]>0,
	\end{equation}
	and forth equality in \eqref{ineq} gives
	\begin{equation}\label{V2}
	V''(\phi_m)=-\frac{b n^2}{c_V^2} \left(\frac{a +c_V/n}{ b }\right)^3
	\left[ -\frac{b}{2 (a  +c_V/n)  n (n-1)} \right]^{n/2} \equiv m^2 > 0 \quad \mbox{i.e.} \quad b >0.
	\end{equation}
	As the result, the parameters of action must satisfy the conditions \begin{equation} \label{cond23}
	b>0, c_V<0, a+c_V/n<0. \end{equation}
	
	Now we are ready to consider the inflationary stage in the framework of our model.
	
	\section{Slow rolling and all that}
	
	First of all, we have to modify our action \eqref{Lscalar} applying definition \eqref{fR} to it.
	The final form of the action is
	\begin{eqnarray}
	\label{l-o}
	S =\frac{m_4^2}{2} \int d^4x \sqrt{-g_4}\mbox{sign}(2a\phi+b)\biggr\{R + K(\phi)\phi_{;\rho}\phi^{;\rho} - 2V(\phi)\biggl\},
	\end{eqnarray}
	where
	\begin{equation}\label{m4-1}
	m_4=\sqrt{V_n}=\sqrt{\frac{2\pi^{\frac{n+1}{2}}}{\Gamma(\frac{n+1}{2})}},\quad m_D=1
	\end{equation}
	with the definitions
	\begin{eqnarray}
	K(\phi) &=& \frac{1}{\phi^2(2a\phi + b)^2}\biggr\{\biggr[(6 - n + \frac{n^2}{2})a^2 + 2(c_1 + c_2)a\biggr]\phi^2 \nonumber\\
	&+& \biggr[\frac{n^2}{2}ab
	+ (c_1 + c_2)b\biggl]\phi + \frac{n(n+ 2)}{8}b^2\biggl\},\\
	V(\phi) &=& - \frac{\mbox{sign}(2a\phi + b)}{2(2a\phi + b)^2}\biggr[\frac{\phi}{n(n - 1)}\biggl]^\frac{n}{2}\biggr\{\biggr(a + \frac{c_V}{n}\biggl)\phi^2 + b\phi + c\biggl\},
	\end{eqnarray}
	and
	\begin{eqnarray}
	c_V &=& c_ 1 + 2\frac{c_2}{n - 1}.
	\end{eqnarray}
	
	There are two main inflationary parameters for the model
	\begin{eqnarray}\label{Lpsi}
	{\cal \bar L} = \sqrt{-g}\biggr\{R + \psi_{;\rho}\psi^{;\rho} - 2V(\psi)\biggl\}.
	\end{eqnarray}
	with the standard form of the kinetic term. There is the scalar spectral index $n_s$ and the scalar/tensorial relative amplitude $r$, besides the running of the spectral index, which are given, in terms of the slow roll parameters as
	\begin{eqnarray}\label{nsr}
	n_s &=& 1 - 6\epsilon + 2\eta,  \\
	%n_T &=& - 2\epsilon,\\
	r &=& 16\epsilon, \label{nsr2} \\
	\frac{d n_s}{d\ln k} &=& - 16\epsilon\eta + 24\epsilon^2 + 2\xi^2. \label{nsr3}
	\end{eqnarray}
	$k$ is the wavenumber.
	
	The slow roll parameters are defined as,
	\begin{eqnarray}
	\label{sr1}
	\epsilon &=& \frac{1}{2}\biggr(\frac{V'}{V}\biggl)^2, \\
	\label{sr2}
	\eta &=& \frac{V''}{V},\\
	\label{sr3}
	\xi^2 &=& \frac{V' V^{'''}}{V^2}.
	\end{eqnarray}
	The primes are derivatives with respect to $\psi$. The parameter $\xi$ is connected with the derivatives of the more usual parameters $\epsilon$ and $\eta$ as it will be seen later. These formulas are derived keeping in mind Lagrangian \eqref{Lpsi} and should be adapted for our action \eqref{Lscalar} with $K(\phi)\neq 1$. It can be done by change of variable according to the equation
	\begin{eqnarray}
	\frac{d\phi}{d\psi} = \frac{1}{\sqrt{K}}
	\end{eqnarray}
	that is connected two scalar field $\phi$ and $\psi$.
	The slow roll parameters \eqref{sr1}, \eqref{sr2}, \eqref{sr3} acquire the form
	\begin{eqnarray} \label{eps}
	\epsilon &=& \frac{1}{2} \left( \frac{V_{,\psi}}{V} \right)^2= \frac{1}{2 K(\phi)} \left( \frac{V_{,\phi}}{V} \right)^2,\\
	\label{eta}
	\eta &=& \frac{ V_{,\psi \psi}}{V(\phi)} = \frac{1}{V(\phi)\sqrt{K}} \frac{\partial}{\partial \phi} \left( \frac{V_{,\phi}}{\sqrt{K}} \right),\\
	\label{xi}
	\xi^2 &=& \frac{V_\psi V_{\psi\psi\psi}}{V^2} = \frac{V_\phi}{V(\phi)^2 K(\phi)}\frac{\partial }{\partial \phi}\biggr\{\frac{1}{K(\phi)}\biggr(V_{(\phi\phi)} - \frac{1}{2}\frac{K_{\phi}}{K}V_{\phi}\biggl)\biggl\}.
	\end{eqnarray}
	for Lagrangian (\ref{l-o}).
	
	During inflation, the two slow-roll parameters must be small:
	\begin{equation} \label{sr4}
	|\varepsilon| \ll 1, \quad |\eta| \ll 1. %,{\color{red} \quad \xi^2 \ll 1????}.
	\end{equation}
	The amount of inflation is given by the
	number of e-folds of the scale factor
	\begin{equation} \label{N}
	N \simeq -\int_{\phi_{in}}^{\phi_{f}} \frac{K V}{V_{,\phi}} d \phi.
	\end{equation}
	valid for the slow-roll regime.
	
	Now we have all the necessary formulas to adjust the parameters of our model.
	
	\section{The parameters fixing}\label{secpar}
	
	In this section, we apply the results obtained above to build the inflationary model \eqref{l-o} which does not contradict the observable values of the spectral index and the tensor-to-scalar ratio  \eqref{nsr}.

\begin{figure}
		\includegraphics[width=7cm]{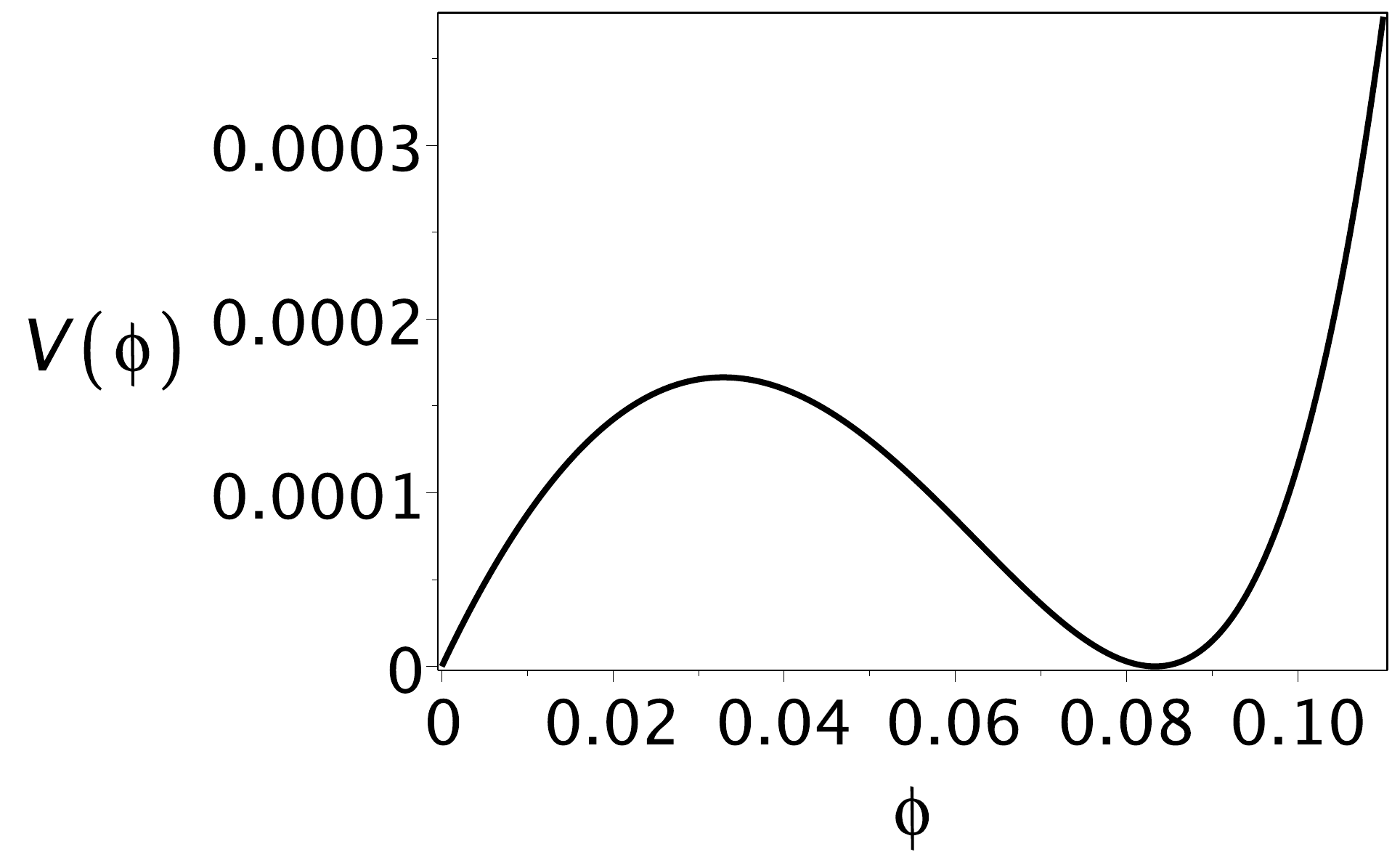} \qquad \includegraphics[width=7cm]{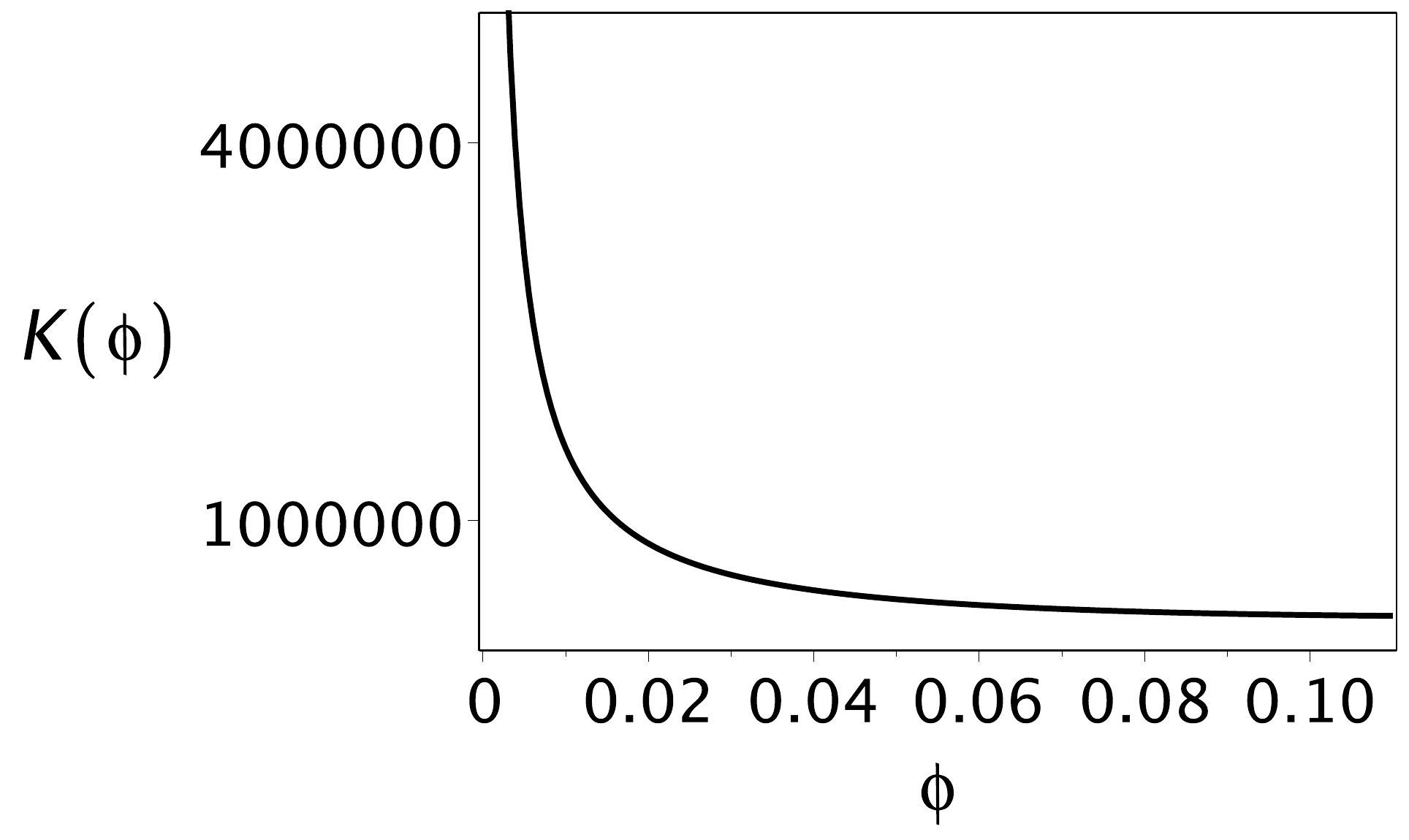}
		\caption{The form of the potential (left) and kinetic term (right) for the parameters $n = 2, b = 1, a = -2, c_V=-8, c_K = 15000.$ The potential minimum is in the point $\phi_m  \simeq  0.083$, $m_D=1.$}\label{VK}
	\end{figure}

	Appropriate parameters
	\begin{eqnarray}\label{param}
	&&n = 2, b = 1, a = -2, c_V=-8, c_K=15000,  \\
	&&\left(c_V = c_ 1 + 2\frac{c_2}{n - 1}, c_K=c_1+c_2. \right) \nonumber %c_1 = -\frac{c_V (1- n)+2 c_k}{n-3}, c_2 = \frac{(c_V -c_k)(1- n)}{n-3},
	\end{eqnarray}
	were found after some selection.
	The combinations $c_V$ and $c_K$ are included in the functions $V$ and $K$ separately which strongly facilitate the selection of the parameters.
	The parameter $"c"$ can be obtained from expression \eqref{cond2}. Restrictions \eqref{ineq} - \eqref{V2} are also taken into account. The potential and the kinetic term of the model are represented in FIG. \ref{VK}.
	
\begin{figure}[ht!]
		\begin{overpic}[width=6cm, height=4cm]{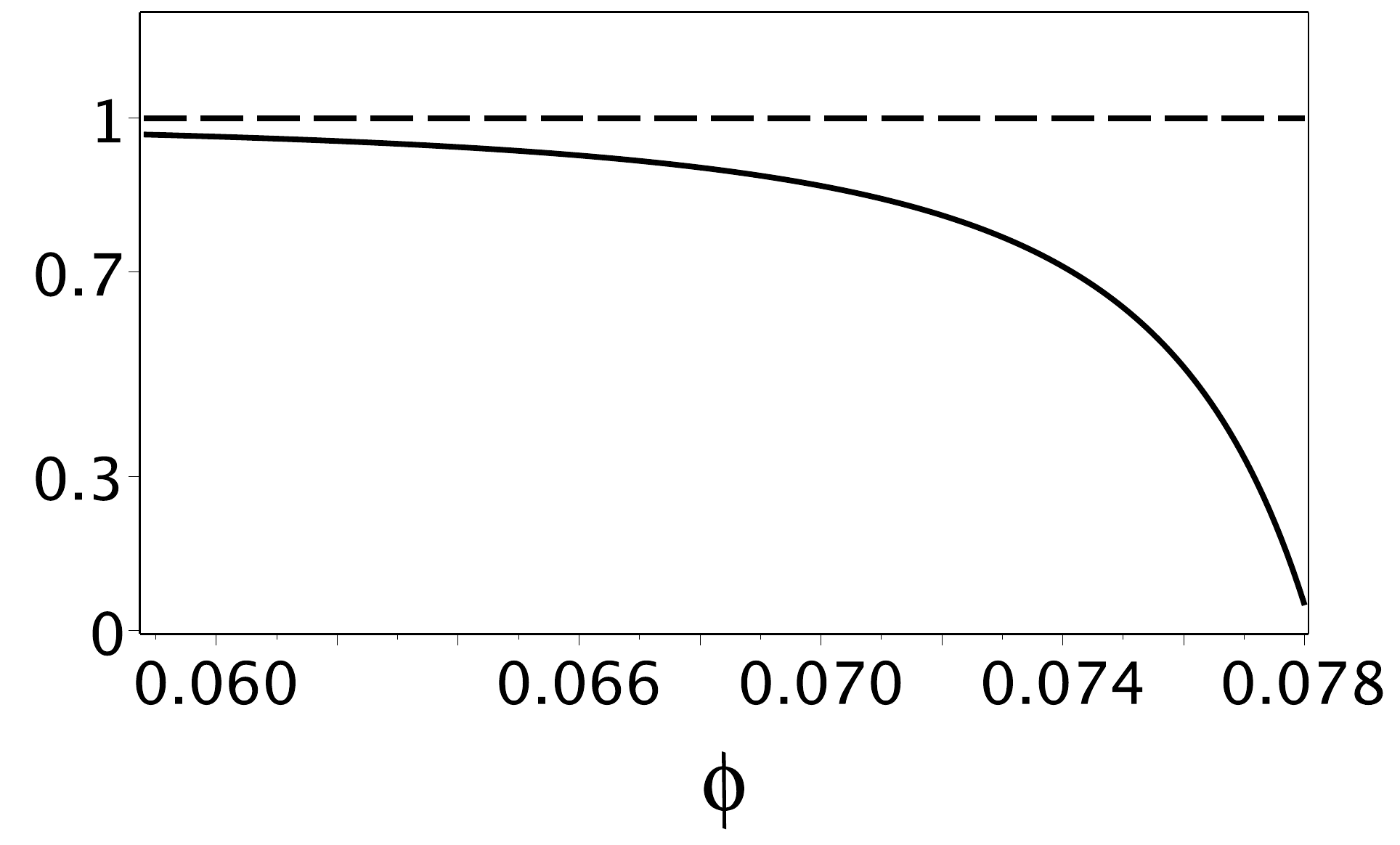} \put(-10,40){   \textrm{ \bf $n_s$}} \end{overpic}
		\hskip1cm
		\begin{overpic}[width=6cm, height=4cm]{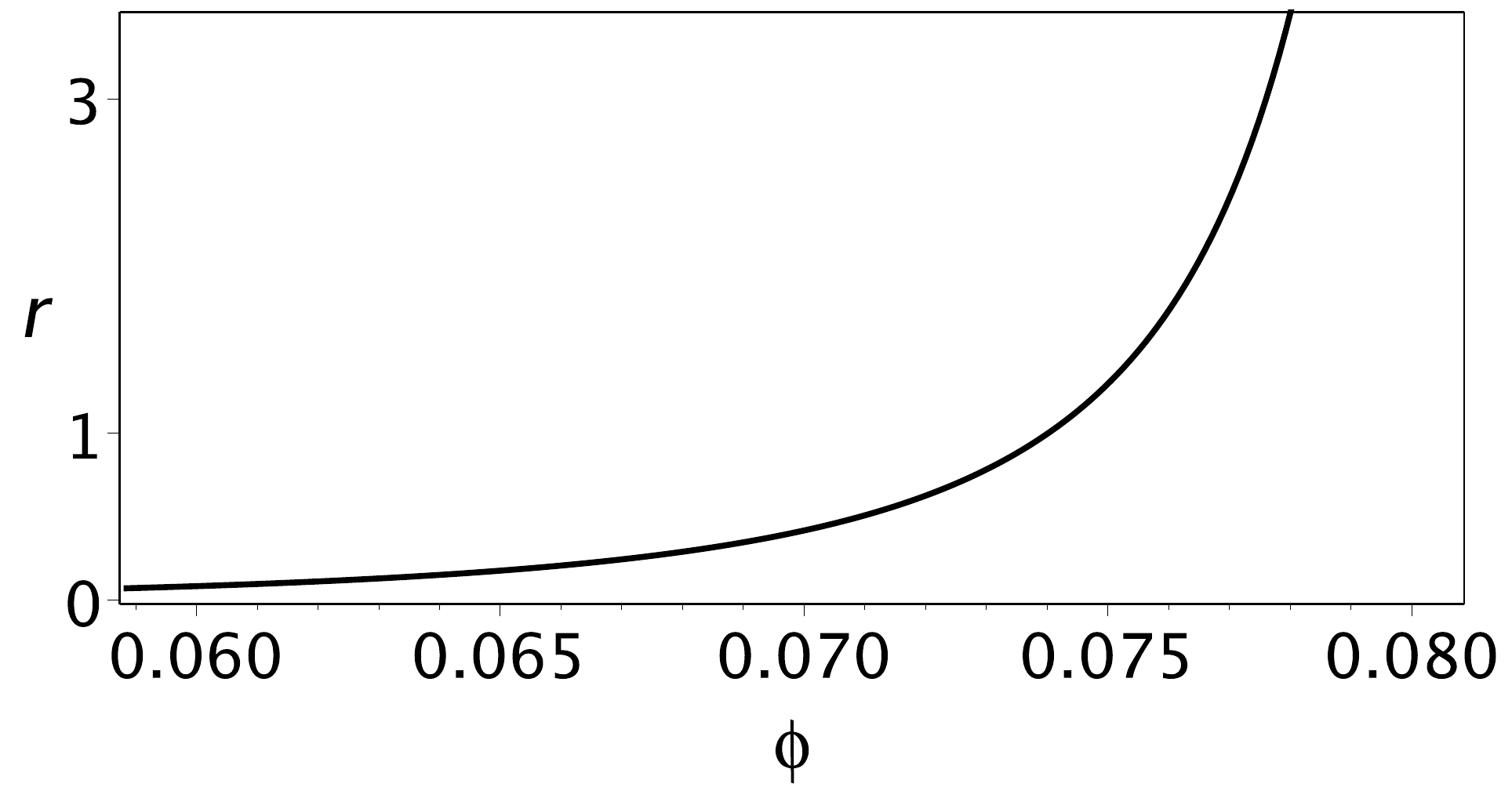} \end{overpic}
		\caption{$n = 2, b = 1, a = -2, c_V=-8, c_K=15000, \phi_m \simeq 0.083 $. Smaller field values relates to larger cosmological scale. The observable values \eqref{nsrObserv} are found for $\phi \sim 0.06.$}
		\label{nsrfig}
	\end{figure}

	The present horizon could be nucleated at any slope of the potential. We suppose that it happens at the left of the potential minimum $\phi_m \simeq 0.083$. The calculations indicate that the initial field value $\phi_{in} \simeq 0.0588$ where slow role parameters are small  is appropriate starting point for the inflation. Final field value $\phi_{f} \simeq 0.0808$ for which the inflation is finished $\varepsilon(\phi_{f} \simeq 0.0808) \lesssim 1, \ \eta(\phi_{f} \simeq 0.0808) \lesssim 1$, are related to the initial field value by the formula \eqref{N} where the e-folds number is assumed to be $N \simeq 60$. Inserting the value  $\phi_{in} \simeq  0.0588$ into \eqref{eps}, \eqref{eta} we come to our the prediction for the spectral index \eqref{nsr}, the tensor-to-scalar ratio \eqref{nsr2} and the running spectral index \eqref{nsr3}
	\begin{equation}\label{result}
	n_s \simeq 0.968,\quad r \simeq 0.07,\quad \frac{d n_s}{d\ln k} \simeq 0.00051
	\end{equation}
	for large scale fluctuations. According to \cite{Cline:2018fuq}, observable values are as follows
	\begin{equation}\label{nsrObserv}
	n_s\simeq0.968,\quad r<0.1,\quad \frac{d n_s}{d\ln k}= -0.0045\pm 0.0067 ,
	\end{equation}
	so that the inflationary model discussed in this paper is valid. The time dependence of $n_s$ and $r$ can be seen from Fig.\ref{nsrfig}. The field evolves starting from the initial value $\phi_{in} \simeq  0.0588$ to the final one $\phi_{f} \simeq 0.0808$.
	
	\begin{figure}[ht!]
		\begin{overpic}[width=6cm, height=4cm]{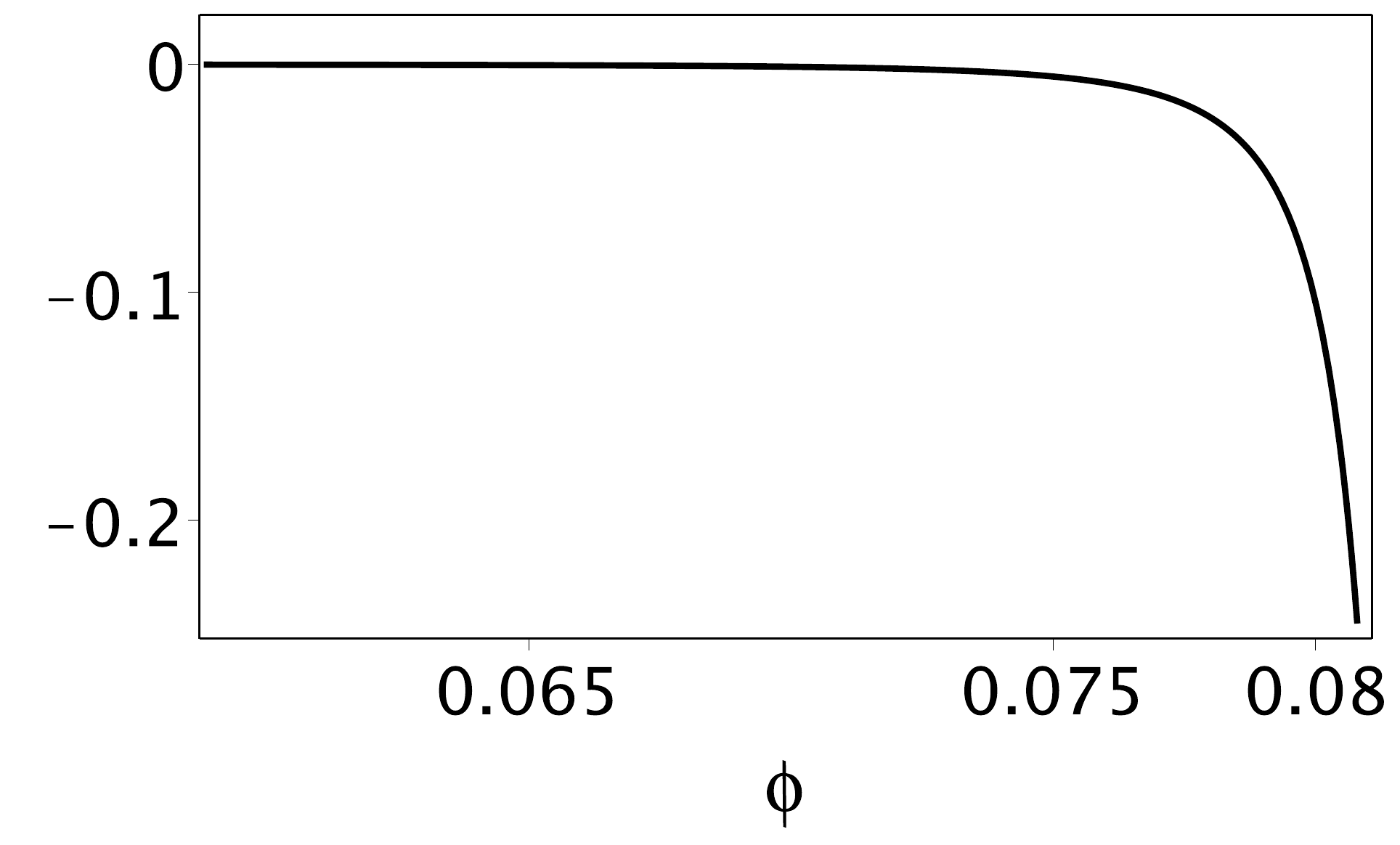} \put(-20,35){   \textrm{ \bf $\frac{\displaystyle d n_s}{\displaystyle d\ln k}$}} \end{overpic}
		\caption{$n = 2, b = 1, a = -2, c_V=-8, c_K=15000, \phi_m \simeq 0.083 $.}
		\label{dns}
	\end{figure}

	%\begin{figure}[ht!] \label{nsr_2}
	%\hskip-16mm        \includegraphics[width=18cm, height=6cm]{nsr_2.jpg}
	%\caption{}
	%\end{figure}
	
	The plot of the potential represented in Fig. \ref{VK} can be used to check inequality \eqref{ineq0} to be sure in the self-consistency of our model. Indeed, according to the plot, $V\simeq 5\cdot 10^{-5}, R_2=\phi\simeq 0.07$ in the middle of inflation. Hence,
	$R_4\simeq 10H^2\simeq 10^2 V\simeq 5\cdot 10^{-3}$ and $R_2\simeq 7\cdot 10^{-2}$ and condition  \eqref{ineq0} holds.
	
	In this paper, we express all numbers in units $m_D=1$ with the effective Planck mass $m_4$ defined by \eqref{m4}. After the inflation is finished and the scalar field has been settled in the minimum of its potential one should restore more physical units valid for the Jordan frame, i.e. the Planck units using the relation \eqref{MPl}. For chosen parameter values \eqref{param} the relation
	\begin{equation}
	M_{Pl}=\sqrt{V_{n}e^{n\beta_m}f'(\phi_m)}m_D
	\end{equation}
	can be used to obtain the value of the D-dim Planck mass.
	For 2-dim extra space ($n=2,\quad V_2 = 4\pi$)
	\begin{equation}
	M_{Pl}=\sqrt{V_{2}e^{2\beta_m}f'(\phi_m)}m_D = \sqrt{8\pi (2a+\frac{1}{\phi_m})}m_D \sim 10 m_D
	\end{equation}
	that means that the D-dim Planck mass $m_D$ is in the order of magnitude  smaller than the Planck mass in the Jordan frame.
	
	\section{Conclusion}
	
	The inflationary model was studied in the framework of non-linear multidimensional gravity without the matter fields. The Ricci scalar of the extra space plays the role of inflation. Parameters of the model are adjusted to satisfy observational values for the spectral index and the tensor-to-scalar ratio. The model has a large parameter ($c_K\sim 10^5$) which is a common feature of the inflationary models.  Our study points that the Kretschmann and the Ricci tensor square terms dominate during inflation so that our model differs essentially from the Starobinsky $R^2$ model \cite{Starobinsky:1980te} leading, however, to similar results. According to the observational data, inflationary models with a single scalar field and simple form of potential are not very perspective. Our model leads to a complex form for the effective potential and kinetic factor that give promising results. We have given an example where the model studied here is in agreement with the observational constraint. A more complete analysis of the space parameter is, of course, necessary.
	
	To construct a more realistic model, it will be interesting to study the behaviour of the extra space during the radiation dominated and the matter-dominated phases. We hope to do this in future work.
	
	\section{Acknowledgement}
	We thank O.F. Piattella for his comments on the final version of the text.
	The work was supported by the Ministry of Education and Science of the Russian Federation, MEPhI Academic Excellence Project (contract N~02.a03.21.0005, 27.08.2013). 
	The work of S.G.R. was also supported by the Ministry of Education and Science of the Russian Federation, Project N~3.4970.2017/BY.
	The work of S.G.R. and  A.A.P. is performed according to the Russian Government Program of Competitive Growth of Kazan Federal University. The work of A.A.P. was also supported by the Russian Foundation for Basic Research Grant N 19-02-00496. J.C.F. is partially supported by {\it Fapes} (Brazil) and {\it CNPq} (Brazil).
	
	%\bibliographystyle{unsrt}
	%\bibliography{Ru-Article}

\end{document}